# Functionalized graphene as a model system for the two-dimensional metal-insulator transition

M. S. Osofsky[1], S. C. Hernández[1], A. Nath[2], V. D. Wheeler[1], S. Walton[1], C. M. Krowne[1], and D. K. Gaskill[1]

[1]Naval Research Laboratory, Washington, DC 20375
[2]George Mason University, Fairfax, VA 22030

Reports of metallic behavior in two-dimensional (2D) systems such as high mobility metal-oxide field effect transistors, insulating oxide interfaces, graphene, and $MoS_2$ have challenged the well-known prediction of Abrahams, *et al*. that all 2D systems must be insulating. The existence of a metallic state for such a wide range of 2D systems thus reveals a wide gap in our understanding of 2D transport that has become more important as research in 2D systems expands. A key to understanding the 2D metallic state is the metal-insulator transition (MIT). In this report, we demonstrate the existence of a disorder induced MIT in functionalized graphene, a model 2D system. Magneto-transport measurements show that weak-localization overwhelmingly drives the transition, in contradiction to theoretical assumptions that enhanced electron-electron interactions dominate. These results provide the first detailed picture of the nature of the transition from the metallic to insulating states of a 2D system.

The excitement generated by the achievement of metallic single layer graphene has obscured the fact that seminal theoretical work predicted that purely two-dimensional (2D) systems should not be metallic [1]. A possible explanation for the metallic behavior in graphene is that massless Dirac electrons exhibit Klein tunneling and are thus, immune to the effects of disorder [2,3] This argument is contradicted by reports that the carriers often have mass [4,5,6], possibly due to disorder and/or the underlying substrate breaking lattice symmetry or the fact that the Fermi energy is far from the Dirac point [2]. Thus, graphene should be described by the theory presented in reference 1 if there is disorder in the potential binding the electrons. The situation is confounded further by later theoretical work showing that Dirac Fermionic systems with no spin-orbit interactions and Gaussian correlated disorder exhibit scaling behavior but should always be metallic [3]. The observed metallic behavior is an unquestionable addition to a series of systems such as high mobility metal-oxide field effect transistors (HMFET) [7] and interface oxides [8] that have demonstrated a 2D metallic state (although the nature of that state for the HMFET's is not well understood). These systems are presumed to be 2D due to their geometry but might have some three-dimensional character since the charge regions extend over finite distances that could explain their metallic transport properties. In contrast, graphene is a model system for studying the 2D metal-insulator transition (MIT) as it is a pure 2D system like $MoS_2$ (which has recently been shown to also have an MIT [9,10,11,12]). In this work, we increase the resistivity of epitaxial graphene through surface functionalization by exposure to low energy plasmas. These results reveal the existence of a 2D MIT in epitaxial graphene where the pre-functionalization values of carrier concentrations and mobilities are $\sim 10^{12}$-$10^{13}$ cm$^{-2}$ and $\sim$700-900 cm$^2$V$^{-1}$s$^{-1}$, values that are out of the range of applicability for the models developed to describe the previous results on the HMFETs [13,14] where the disorder is thought to be screened by high mobility electrons. It is possible that a more recent general scaling model that was developed for the high mobility case, and allows for the existence of a 2D MIT [15], can be used to model the graphene system as well. The results presented here demonstrate that the strongly localized state is separated from the metallic state by a weakly localized phase with conductivity, σ, ~log(T) similar to results recently reported for thin films of $RuO_2$ [16].

Previous work has shown that an MIT does indeed exist for graphene: it is well established that graphene can have a metallic state and Chen, *et al.* have demonstrated that insulating samples result through exposure to ion damage [17]. Furthermore, Bostwick, *et al.* observed an MIT by showing a large increase in room temperature resistance accompanied by a breakdown of the quasi-particle description as determined from photoemission in graphene exposed to atomic hydrogen although no low temperature data were reported [ 18 ]. Key to understanding the 2D MIT is the study of metallic transport near the transition. In 3D materials, it is known that weak-localization (WL) and enhanced electron-electron interactions (EEI) control the metallic transport properties near the MIT. For metallic graphene with moderately high mobility, there have been several studies reporting WL and/or EEI [19,20,21,22,23,24,25,26,27]. Those results, while suggestive, are for graphene relatively far from the MIT where WL and EEI can be treated as corrections to the conductivity. That approach fails near the MIT, a quantum phase transition, where scaling models of phase transitions are needed to describe the properties [28,29]. Thus, it appears that a scenario analogous to the three-dimensional case where the disorder driven MIT is described by a phase diagram with four regions [30,31]: insulating, critical, amorphous metal, and conventional metal can be observed for the 2D case. In the present study the systematic increase in the graphene's sheet resistance resulting from exposure to low energy plasmas has been used to determine a critical exponent of this phase transition and estimate the relative contributions of WL and EEI as the strongly localized phase is approached.

**Preparing and functionalizing epitaxial graphene**

Several samples of epitaxial graphene were grown via Si sublimation from nominally on-axis SiC (0001) substrates (32). Prior to graphene growth, substrates were etched in hydrogen at 1520°C, 100 mbar for 10-30 min. to remove polishing damage. Graphene was then synthesized in 10 standard liters per minute of Ar at 1540°C, 100 mbar for 25-35 min. These conditions resulted in graphene with an average thickness of 1.5 layers as determined by x-ray photoelectron spectroscopy. The samples were then fashioned into a pattern that enabled standard four-probe resistivity and Hall measurements (see supplemental materials). Each sample was then systematically exposed to electron beam generated plasmas produced in mixtures of $O_2$, $SF_6$, or $N_2$ to introduce oxygen-, fluorine-, or nitrogen-functional groups [33,34]. Some samples were also selectively exposed to a vacuum anneal after plasma treatments to reduce the resistance. Increasing dosage is indicated by an increasing numerical symbol, i.e., N0 (Nitrogen series, no dose), N1, N2, etc.; see Table S1 for details. Raman measurements indicated that the graphene signature was present after functionalization (see supplemental materials).

As grown, the samples had resistance, R, ~1000Ω/□, carrier concentrations of ~$10^{12}$-$10^{13}$ cm$^{-2}$, and mobilities ~700-900 cm$^2$V$^{-1}$s$^{-1}$ measured at room temperature (see supplemental materials). It is important to note that very low currents were used for the transport measurements to ensure that local heating did not obscure the results at low temperature (see supplemental materials). The carrier concentrations are comparable to those reported for HMFET's (~$10^{10}$-$10^{12}$ cm$^{-2}$) [35] with the starting mobility values higher than those reported for oxide interface FET systems [36,37,38], and smaller than those reported for conventional HMFET devices, ~$10^4$ cm$^2$V$^{-1}$s$^{-1}$ [35]. By exposing the graphene to the plasmas, the room temperature resistance eventually increased to values that exceed the quantum resistance, $h/(2e^2)$. The samples can thus be driven through the MIT by systematically exposing the graphene to increasing plasma doses and vacuum anneals (Figure 1). The amount of induced disorder varied by element with F and O having the strongest influence (see supplemental materials).

**Transport properties: 2D metal-insulator transition**

Figures 1a and b show data for N and O exposures clearly demonstrating transitions from conventional metallic behavior, dR/dT>0, for dose 0 and 1 (with low temperature deviations) to a state with dR/dT<0 for higher doses. At the highest oxygen doses the graphene exhibits behavior consistent with 2D variable range hopping (VRH), R~exp(1/T$^{1/3}$), demonstrating the transition to a strongly localized insulating state (inset fig 1b). The data for lower O exposures and for the N exposures did not show VRH behavior. Fluorine was so effective in increasing the resistance that only a few R(T) curves could be obtained before the strongly localized state was achieved.

The original theoretical work on WL that described the three dimensional MIT [1] also predicted that all 2D systems will be insulators with σ~log(T). Later work indicated that this log(T) behavior would also result from enhanced electron-electron interactions in diffusive 2D systems [39]. Indeed, work on Si MOSFETS [40,41] and very thin films demonstrated this log(T) behavior [42,43,44,45,46]. Figure 2 shows plots of conductance as a function of log(T) for the data in Figure 1. These curves clearly show these samples having 2D transport characteristics at low temperatures. The relevant low temperature data were fit to

$$\sigma = \sigma_{1K} + \sigma_2 \log(T). \quad (1)$$

Those fits are shown as solid lines in Figure 2.

One of the key issues in understanding the MIT is the slope of the critical phase line, or "mobility edge," that describes the transition into the strongly localized state. In three dimensions, this line is usually defined as the relationship between a driving parameter, generically labeled as p, and $\sigma_0^{3d}$, the value of conductivity extrapolated to T=0 [28,29,30,42,47] The usual formulation is $\sigma_0^{3d}$~(p-p$_c$)$^\zeta$ where p$_c$ is the critical value of p (where $\sigma_0^{3d}$=0) and where ζ is a critical exponent [28,29,30,48]. Experimentally, p is often the carrier concentration. Another choice for p is the bare, high-energy conductivity that can be approximated by the room temperature conductivity [47]. It has been shown that in three dimensions ζ=1/2 in Si:P [49,50] while ζ=1 in disordered metals [28,29,42,47].

In two dimensions, the data analysis is complicated by the fact that the data cannot be extrapolated to T=0 since σ has a log(T) behavior. In this case, one can obtain an analog of the mobility edge by replacing $\sigma_0^{3d}$ with σ$_{1K}$ from equation 1 so that the relevant expression becomes σ$_{1K}$~(σ$_{300K}$-σ$_c$)$^\zeta$ where σ$_c$ is the value of σ$_{300K}$ for which σ$_{1K}$=0. If the phase transition is governed by a scaling law, this formulation should capture the nature of the transition (*i.e.* whether it is continuous and, if so, the value of ζ). This mobility edge is plotted in figure 3 using data from the three types of exposures. The plot clearly shows that the transition is continuous with ζ=1, similar to many disordered 3-d systems.

**Weak-localization vs. enhance electron-electron interactions**

There have been several theoretical approaches for describing diffusive transport in disordered 2D conductors using scaling [28,51,52,53,54]. These models predict insulating behavior in 2D. However, they require the suppression of WL, either by strong spin-flip scattering, strong spin-orbit coupling, or a strong internal magnetic field (e.g., in a ferromagnet) leaving EEI as the relevant phenomenon near the MIT.

Magneto-transport measurements, including the Hall resistance, provide a means to distinguish between the contributions of WL and EEI to conductance. Previous work on samples far from the insulating phase has demonstrated a wide variety of behaviors that include WL and/or EEI corrections to the conductance [19,20,21,22,23,24,25,26,27]. None of those studies considered samples close to the strongly localized state.

For our functionalized samples the Hall resistance, $R_{Hall}$ showed a log(T) temperature dependence at low temperatures (see supplemental materials). This behavior is consistent with that described by Altshuler and Aronov for EEI in disordered systems [39]

$$\frac{\Delta R_{Hall}}{R_{Hall}} = \gamma \frac{\Delta R}{R} \qquad (2)$$

where $R_{Hall}$ is the Hall resistance, R is the resistance, and γ=0 for no EEI [55], γ=2 for EEI, and γ>2 if there is spin-orbit interaction. The values of γ at 1.75K were determined from the slopes of the R(T) and $R_{Hall}$(T) data below 10K. These results are plotted in figure 4 as a function of $\sigma_{1K}$, a measure of the distance to the exponentially localized state. It is apparent that γ<2 with a clear trend in which γ approaches 0.2 as the system approaches the exponentially localized state. This is in contrast to the results reported by Lara-Avila, *et al*. [19] who found γ≥2. The source of this discrepancy may be the fact that the mobilities of the samples studied in ref. 19 were ~6000-7000 cm$^2$/(V-s) which is 10-100 times larger than those measured in this work and indicate the measurements were far from the MIT. The systematic decrease in γ as our system approaches the strongly localized phase is similar to behavior observed in Si MOSFETS where γ~2 for low channel resistance but approached 1 as the channel resistance increased [56]. Thus, the Hall resistance results show that the influence of EEI decreases as the system approaches the strongly localized phase and that transport properties are dominated by WL.

Magneto-resistance (MR) measurements, which are also influenced by WL and EEI, were performed simultaneously to the Hall measurements to further explore how they influence transport. Figure 5 shows the MR results at 1.75K for various plasma exposures plotted in the manner suggested by the theory of McCann, *et al*. [57]. In that theory, the expression for the MR, $\Delta\rho(B) = \rho(B) - \rho(0)$, is

$$\Delta\rho(B) = -\frac{e^2\rho^2}{\pi h}\left[F\left(\frac{B}{B_\varphi}\right) - F\left(\frac{B}{B_\varphi + 2B_i}\right) - 2F\left(\frac{B}{B_\varphi + B_*}\right)\right] \qquad (3)$$

where *F* is a function containing the natural logarithm and the digamma function,

$$F(z) = \ln(z) + \psi\left(\frac{1}{2} + \frac{1}{z}\right), \tag{4}$$

and ρ is the resistivity. Subscripted magnetic fields in equation 3 are simply the effective magnetic representations of the relaxation times,

$$B_\varphi = \frac{\hbar c}{4De\tau_\varphi} \quad ; \quad B_i = \frac{\hbar c}{4De\tau_{inter}} \quad ; \quad B_* = \frac{\hbar c}{4De\tau_*} \tag{5}$$

where $\tau_\varphi$ and $\tau_{inter}$ are the relaxation times for inelastic decoherence and intervalley scattering, respectively, and intravalley scattering and trigonal warping are folded into intervalley scattering through

$$\tau_*^{-1} = \tau_{warp}^{-1} + \tau_{intra}^{-1} + \tau_{inter}^{-1}. \tag{6}$$

We note that the curves in Fig. 5 are for $\rho_{xx}$ data. Formally, the inverse of the conductivity tensor should be used [20] but the contribution of $\rho_{xy}$ is negligible and can be ignored. The plots clearly have the shape and negative MR that is characteristic of WL. In contrast, for EEI the MR is characterized by a $B^2$ magnetic field dependence [20,26,58] and is usually positive [28]. The solid lines in figure 5 are fits to equation 3. The characteristic time scales resulting from the fits are shown in the supplemental materials. While this model appears to provide a good fit to the data, it must be emphasized that the values of the parameters extracted should not be taken too seriously since the theory describes a correction to resistance of a weakly disordered metal, a situation far from that of graphene close to the transition to the strongly localized state that is described here. The crucial finding is that, consistent with the Hall data, the MR results indicate that WL is the dominant transport phenomenon, contradicting the assumptions of the prevailing theoretical treatments of 2D disordered systems that treat EEI as the dominant mechanism influencing transport [28,51,52,53,54].

More recently, Dobrosavljević, *et al.* (15) extended the theory of Abrahams, *et al.* [1] to include electron-electron interactions (for screening but not EEI) by relaxing the assumption that the scaling function is monotonic and negative for "large" conductance. This modification results in the prediction of a 2D MIT. To explore whether this theory can describe our results, we use the scaling model conductance from reference 15:

$$g(\delta n, T) = g_c e^{sgn(\delta n) A \left[\frac{T_0(\delta n)}{T}\right]^{1/(\nu z)}} \tag{7}$$

where $g_c$ is the critical conductivity for the MIT, sgn( ) is the sign operator, $\delta n = (n - n_c)/n_c$ with $n$ and $n_c$ the carrier and critical carrier concentration, $T_0$ is a crossover temperature which has $T_0(\delta n) \sim |\delta n|^{\nu z}$, $A$ is a dimensionless constant of order unity, $\nu$ is the correlation length exponent, and $z$ is the dynamical exponent relating temperature and length scale, $L$, by $T \sim L^{-z}$. By expanding the exponential

to two terms, expanding the natural logarithm, combining terms, and dropping the A factor, we obtain

$$g(\delta n, T) = g_c + g_c \ln\left[1 + \frac{|\delta n| sgn(\delta n)}{T^{1/(vz)}}\right] \quad (8)$$

Near the critical point ($\delta n = 0$), and neglecting the sgn operator for the moment, generates the following formula:

$$g(\delta n, T) = g_c + \ln(|\delta n|) + \frac{1}{v_s z}\ln(T) \quad (9)$$

which is consistent with the weak localization approach discussed earlier, and the use of equation [1] to fit our data. The relationship between $v$ and $v_s$ is $v = v_s \,\text{sgn}\left[-d\beta(t)/dt\big|_{g=g_c}\right]$, $v^{-1} = \left|d\beta(t)/dt\big|_{g=g_c}\right|$, with $v>0$, where $\beta$ is the scaling function from reference 15. Since $\beta$ is non-monotonic, its derivative would have to have the sign that results in $v_s > 0$ for our case.

While this model appears to describe the graphene results, it must be noted that the original motivation for this theory was the discovery of an apparent 2D MIT in HMFET's. It isn't clear that the theory is applicable here, especially since it also predicts that these 2D systems are perfect metals that are not Fermi liquids in the metallic state, a description that does not apply to graphene. It is possible that there is a rich phase space that encompasses both the low and high mobility cases that should be pursued further although it is not clear that this model can be used to quantitatively analyze the data. Such an analysis will probably require a more detailed two-parameter scaling model that explicitly includes EEI and WL.

In conclusion, we have demonstrated the existence of a continuous 2D MIT in epitaxial graphene, a model 2D electronic system. These results contradict a theoretical analysis that predicts robust metallic behavior in graphene and clearly show that the phase diagram is analogous to that for three dimensions with the conductivity having a log(T) temperature dependence rather than $T^{1/2}$ dependence above the strongly localized phase. Magnetoresistance and Hall resistance measurements reveal that WL dominates as the strongly localized state is approached, contrary to the assumptions of renormalization group theories that only treat EEI to describe 2D disordered systems and do not predict a 2D MIT. These results are consistent with a scaling model by Dobrosavljevic', *et al.* that predicts an MIT in two dimensions and suggests that a more complete theory is needed for the 2D MIT.

**Acknowledgements.** This work was supported by the Office of Naval Research. The authors wish to acknowledge Steve Hair for providing data processing software.

**Author contributions.** M. S. O. designed, performed, and analyzed the data, for the transport experiments. V. D. W. and D. K. G conducted growth experiments and produced epitaxial graphene samples used for this experiment. A. N. patterned and optically characterized the samples. S. C. H. functionalized the samples, characterized the devices using XPS and Raman spectroscopy, and analyzed surface spectroscopy data. C. M. K. modeled the transport data. D. K. G. also assisted in analyzing the data. S.G.W. developed the plasma processing system and helped analyze the results. All authors contributed to the manuscript.

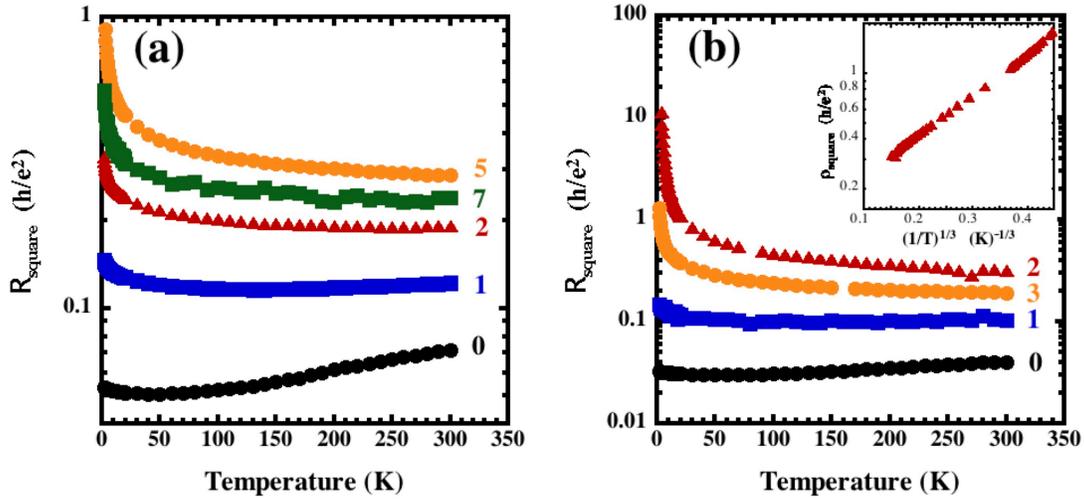

Figure 1. Resistance/square (plotted in units of $h/e^2$) for graphene exposed to (a) nitrogen- and (b) oxygen-containing plasmas. Estimates for the total ionized species produced for each sample are shown in Table S1. Inset (b): Log(resistance/square) vs. $(1/T)^{1/3}$, the behavior expected for 2D variable range hopping, for oxygen sample 2. The curves denoted by "0" are for untreated graphene while the curves denoted "1-7" correspond to increasing plasma dose, with values given in Table S1 (see supplemental materials).

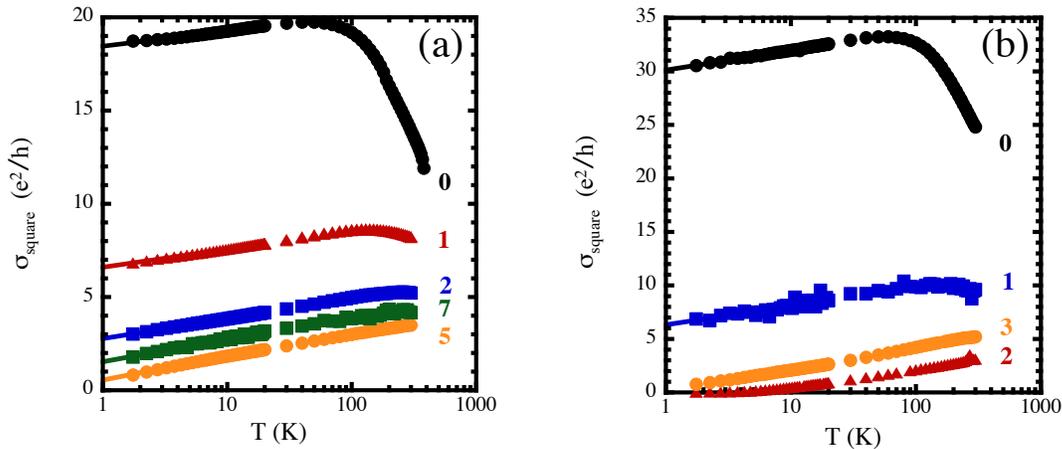

Figure 2. Conductivity per square plotted vs. log(T) as a function of dose for the data in Figure 1 for plasma exposures to (a) nitrogen and (b) oxygen. The solid lines are extrapolated fits to $\sigma=\sigma_0+\sigma_1\log(T)$ for T<10K.

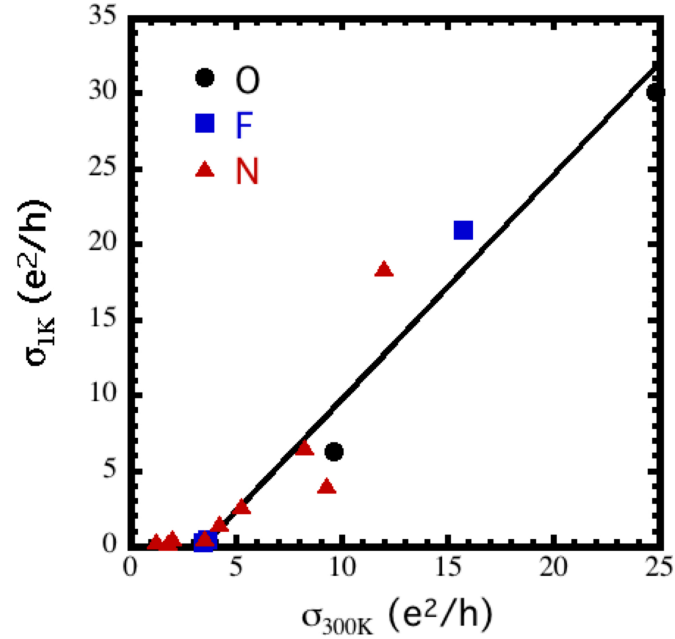

Figure 3. Conductivity at 1K vs. conductivity at 300K showing the continuous nature of the conductivity as the sample approaches the MIT similar to the linear "mobility edge" observed in many three dimensional systems. The line is a linear fit to the data for $\sigma_{1K}>0$. F, O, and N refer to samples subject to plasmas containing $SF_6$, $O_2$, and $N_2$, respectively

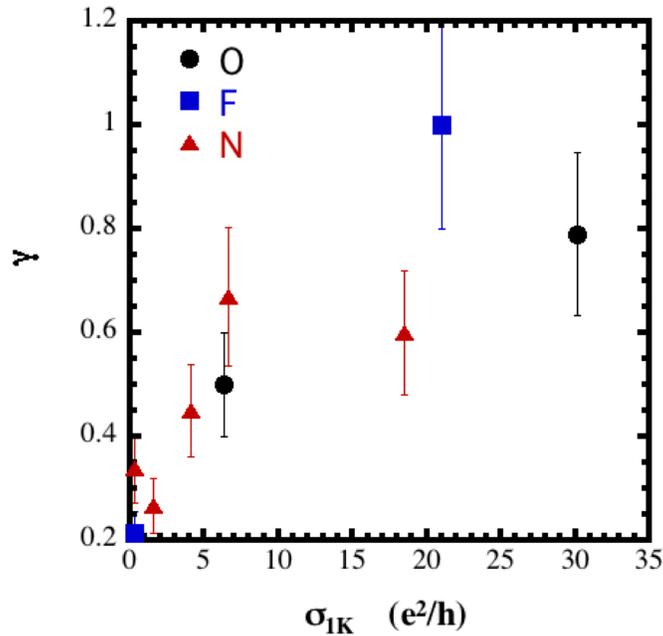

Figure 4. The ratio γ=(ΔR$_{Hall}$/R$_{Hall}$)/(ΔR/R) at 1.75K as a function of σ$_{1K}$, a measure of the distance to the exponentially localized state.

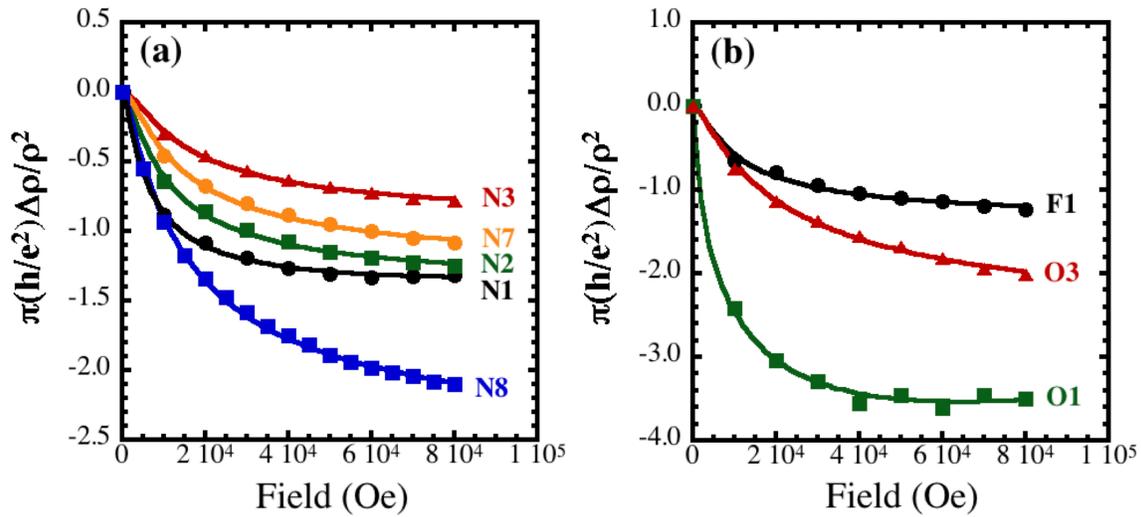

Figure 5. Magneto-resistance data obtained at 1.75K for samples treated in (a) nitrogen- and (b) oxygen- and fluorine-containing plasmas. Exposures and σ(1K) for the doses are listed in Table S1. The solid symbols are experimental data and the solid lines are fits to the weak-localization model of McCann, *et al*. (Ref. 57).

Supplemental Materials

Functionalized graphene as a model system for the two-dimensional metal-insulator transition

M. S. Osofsky[1], S. C. Hernández[1], A. Nath[2], V. D. Wheeler[1], S. Walton[1], C. M. Krowne[1], and D. K. Gaskill[1]

[1]Naval Research Laboratory, Washington, DC 20375
[2]George Mason University, Fairfax, VA 22030

*1.     Graphene Growth.* Epitaxial graphene (EG) samples were synthesized by Si sublimation on the nominally on-axis, (0001) (also called the Si-face) of 8 x 8 mm$^2$ semi insulating (>10$^9$ Ωcm) 6H-SiC substrates (II-VI, Inc.) in a commercial Aixtron VP508 chemical vapor deposition reactor. Prior to growth, samples underwent an *in situ* H$_2$ etch, using palladium purified gas, at 1520°C at 100mbar for 10 to 25 minutes to remove any surface damage thus producing a uniform bilayer stepped surface. Graphene formation followed at a temperature of about 1540°C, in a high purity argon atmosphere at 100 mbar for 25 to 35 minutes [*1*].

*2.     Device Fabrication.* Hall bars were lithographically patterned on the EG by traditional photolithography using LOR and S1811 photoresists to create a clean graphene surface, as described elsewhere[*2*] and this subsequently ensures relatively low metal contact resistance. This process produced atomically smooth samples with clean surfaces (Figs. S1(b)-(d)). The resulting Hall bars were 10 μm wide and 110 μm long, with Ti/Au contacts (10 nm/100 nm). External leads of Au wire connecting the device to a sample holder were added using a K&S Ball Bonder (Model 4522) (Fig S1(a)).

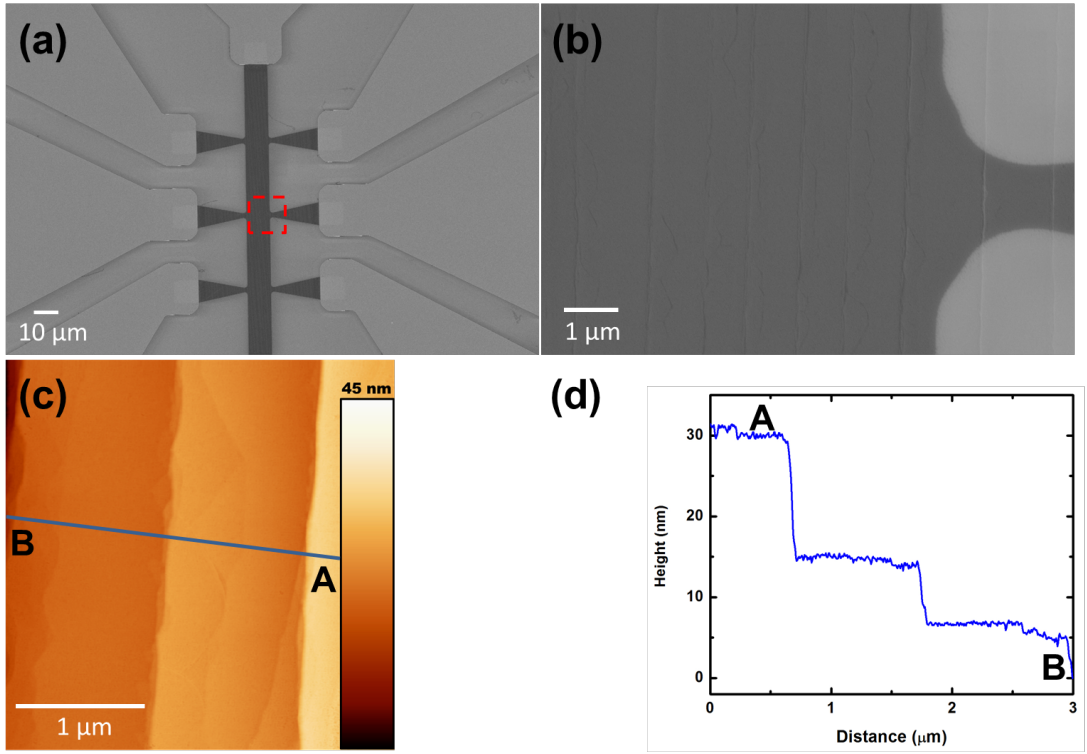

Figure: S1(a) In-lens Scanning Electron Microscope image of a representative Hall structure. (b) Magnified In-Lens SEM image of the boxed region of (a) to demonstrate clean surface after device fabrication. (c) AFM height image of the same device showing the terrace and step structure of the samples. The rms roughness of a 0.5 × 0.5 μm² region on a terrace is about 0.18±0.03 nm, which is similar to the as-grown sample roughness measured in the same fashion. (d) Line scan from (c) showing typical step heights.

3. *Plasma Processing.* While various plasma sources have been used in the synthesis and modification of graphene [*3*], electron beam generated plasmas are well-suited for chemical functionalization, as they are capable of delivering a flux of reactive species while limiting the ion kinetic energies to a few eV [*4*] - a value that is at or near the carbon-carbon bond strength. Thus, they provide the ability to chemically modify graphene without etching or introducing unwanted changes [*5,6*] For this work, pulsed, electron beam driven plasmas were produced in mixtures of $N_2$/Ar, $O_2$/Ar, or $SF_6$/Ar to generate the desired functionalities. High-energy electron beams were created by applying a - 2 kV pulse to a linear hollow cathode for a duration of 2ms at a duty factor of 10%. The emergent beam passed through a slot in a grounded anode and was then terminated at a second grounded anode located further downstream. The electron beam was magnetically confined to minimize spreading via collisions with the background gas, producing a sheet-like plasma. The system base pressure was maintained at ~$1 \times 10^{-6}$ Torr prior to processing by a turbo molecular pump. Reactive gases were introduced at 5% of the total flow rate with argon providing the remainder. The operating pressure (25-90mTorr) was controlled by adjusting the total flow rate (100-180 sccm). The EG samples were placed on a processing stage adjacent to the plasma at a distance of 2.5 cm from the electron-beam axis. All processing experiments were performed at room temperature. For consistency, a single sample was used for each gas mixture and subjected to repeated plasma treatments and measurements. As such, the reported material properties of any one particular exposure (dose) is the culmination of that exposure plus any prior exposures.

4. *Plasma Dose.* It is difficult to precisely know the fluence of reactive species delivered to the graphene surface across a range of operating conditions and background gases. However, since high-energy beam electrons are the primary driver of species production, it is possible to use the processing parameters along with a few assumptions, to estimate the dose of plasma-produced species at the surface [7] and compare the results for the various processing conditions. In particular, a comparison of the total production of the primary ionization product in each background gas for a given set of operating parameters serves as a reasonable proxy for the dose of reactive species. Table S1 shows the relative cumulative dose for samples processed under the conditions listed. The notation F1-3, O1-2, and N1-7 refer to sequential exposure to plasmas produced in backgrounds containing $SF_6$, $O_2$, and $N_2$, respectively. Samples labeled O3, O4 and N8 were subject to a vacuum anneal rather than plasma exposure and since annealing effectively removes functional groups, dose is meaningless and thus is omitted. Samples labeled as N0, O0 and F0 refer to the pristine- unfunctionalized devices.

Table S1. Process conditions and relative dose for the samples studied in this work. F, O, and N refer to operating backgrounds containing $SF_6$, $O_2$, and $N_2$, respectively.

| Reactive Background Gas | Sample | Operating Pressure (mTorr) | Plasma Exposure Time (sec) | Cumulative Dose (a.u.) | Vacuum Anneal |
|---|---|---|---|---|---|
| SF$_6$ | F1 | 50 | 6 | 6.63 | - |
| | F2 | 50 | 6 | 9.34 | - |
| | F3 | 90 | 6 | 13.68 | - |
| O$_2$ | O1 | 50 | 6 | 1.88 | - |
| | O2 | 75 | 6 | 5.90 | - |
| | O3 | - | - | - | 450 °C, 1 hr |
| | O4 | - | - | - | 600 °C, 3 hr |
| N$_2$ | N1 | 90 | 6 | 2.59 | - |
| | N2 | 90 | 6 | 7.10 | - |
| | N3 | 75 | 6 | 9.80 | - |
| | N4 | 75 | 6 | 12.91 | - |
| | N5 | 90 | 6 | 17.41 | - |
| | N6 | 90 | 12 | 27.42 | - |
| | N7 | 90 | 24 | 47.44 | - |
| | N8 | - | - | - | 300 °C, 2 hr |

5. *Surface Characterization.* Ex-situ surface diagnostics were performed before and immediately after each sequential plasma exposure to determine the starting material quality and chemistry and the changes resulting from plasma treatment. Ex-situ x-ray photoelectron spectroscopy was performed using a Thermo Scientific K-Alpha spectrometer with a monochromatic Al-K (1486.6 eV) source. The measurement spot size was 100 μm at 100 scans and 100 ms dwell time. The XPS measurements were performed on the lithographically patterned 200 by 200 μm Hall structure.

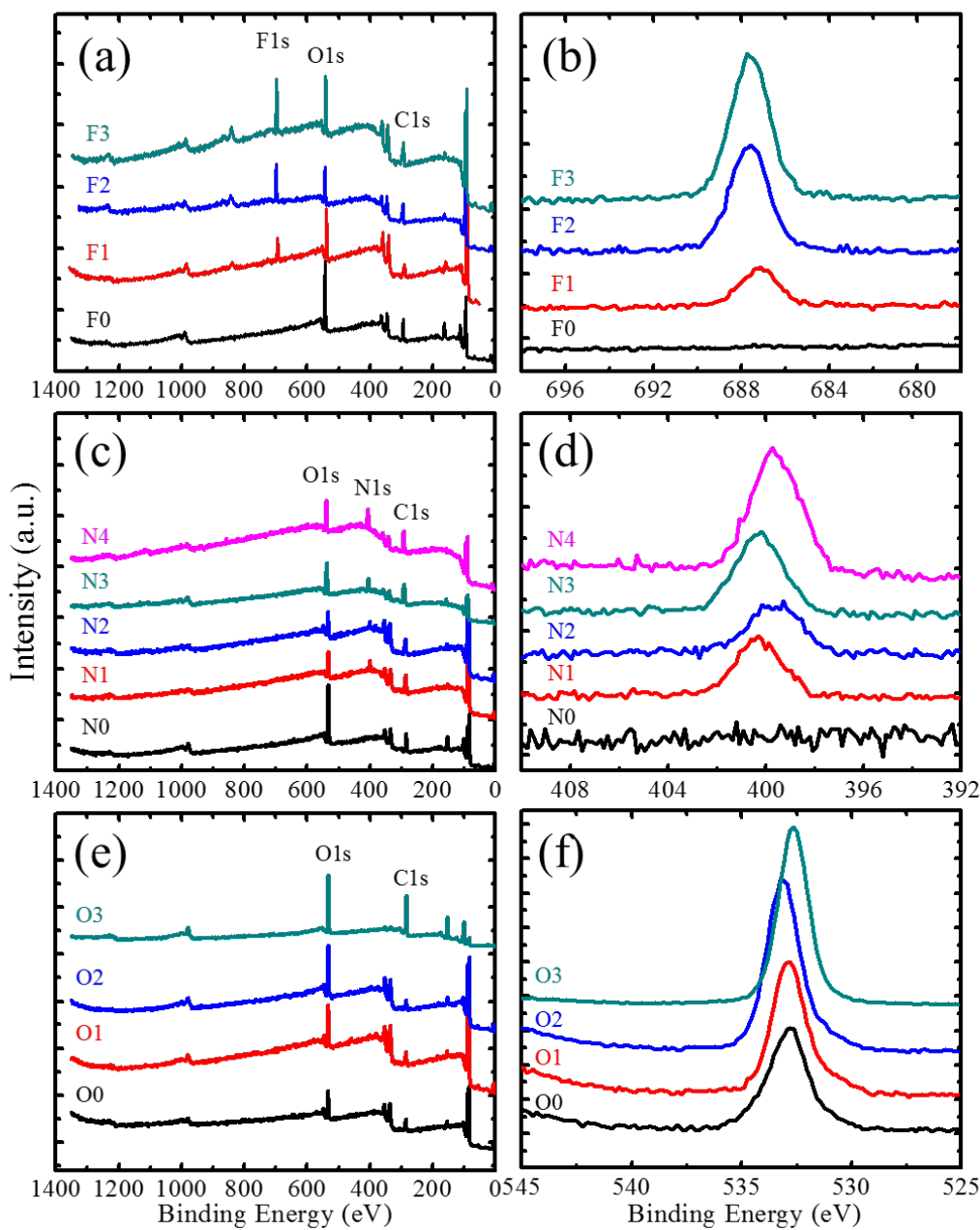

Figure S2: XPS spectra fluorine, nitrogen, and oxygen modified graphene devices. The (a) survey spectra and (b) high resolution F1s spectra for the sequential functionalization of fluorinated devices. (c) Shows the survey and (d) high resolution N1s spectra of the sequential nitrogen functionalization. (e) Shows the survey and (f) and high resolution O1s spectra of oxygen functionalized device.

Chemical changes and the resulting bonding characteristics in the graphene due to plasma processing of individual devices are shown in Figure S2. Following plasma exposures the XPS survey spectra showed a clear introduction of either fluorine, nitrogen or oxygen species as denoted in Figure S2 (a), (c) and (e), respectively. This introduction of chemical species increased with increasing plasma dose. The high-resolution core level spectra shows the general increase in each curve's intensity demonstrating higher amount of chemical species for each type (F, N or O) (Fig. S2 (b), (d) and (f)). Slight shifts in the spectra could be associated with sample charging due to the insulating nature of the SiC and to the limited graphene material (200 µm x 200 µm). Oxygen peaks present on the survey spectra prior to functionalization could be due to small inadvertent sampling outside of the graphene mesa, and likely in the form of $SiO_2$. The high resolution Si 2p spectra shows two peaks; possibly due to Si or SiC and $SiO_2$ at 99-100 eV and 103-104 eV, respectively. Figure S2 (a) and (b) show increasing fluorine content with each sequential dose. Based on the peak positions of the F1s, the fluorine functionalities added were C-CF, C-F and C-$F_2$, with the latter increasing in content at higher fluorine dosages. This is evident from the peak position shifts towards higher binding energies. For the oxygen case; after oxygen plasma functionalization, features on the O 1s spectra arose at three different locations corresponding to (Si-O) bonding at ≈534.3eV, ethers or alcohols (C-O-C, C-O, or C-OH) at ≈533.3eV and carbonyl groups (C=O) at ≈532.2eV. For the nitrogen scenario, the assignment of the components of the nitrogen functionalized epitaxial graphene device at various operating conditions was challenging due to the overlapping binding energies of nitrogen and oxygen species with those of the interfacial layer. However based on the combined features of the C1s and N1s high resolution spectra, the identifiable peaks are Si-C, C-C $sp^2$, and interfacial layer, respectively. The N1s corroborates the presence of nitrogen functionalities present primarily in the amide and pyrrolic configurations. Oxygen peaks present on all the survey spectra prior to functionalization could be due to small inadvertent sampling outside of the graphene mesa, and likely in the form of $SiO_2$. The high resolution Si 2p spectra in figure S3 shows two peaks that can be attributed to Si or SiC and $SiO_2$ at 99-100 eV and 103-104 eV, respectively. The peaks remain even after the cumulative functionalization of the device (fluorine shown here), and presumed to be characteristic of the substrate. Importantly, there is an evident increase in chemical species bound to the surface of the individual device with each additional plasma exposure without removal of the graphene back bone. These chemical species are covalently attached to the graphene carbon back bone generating surface defects.

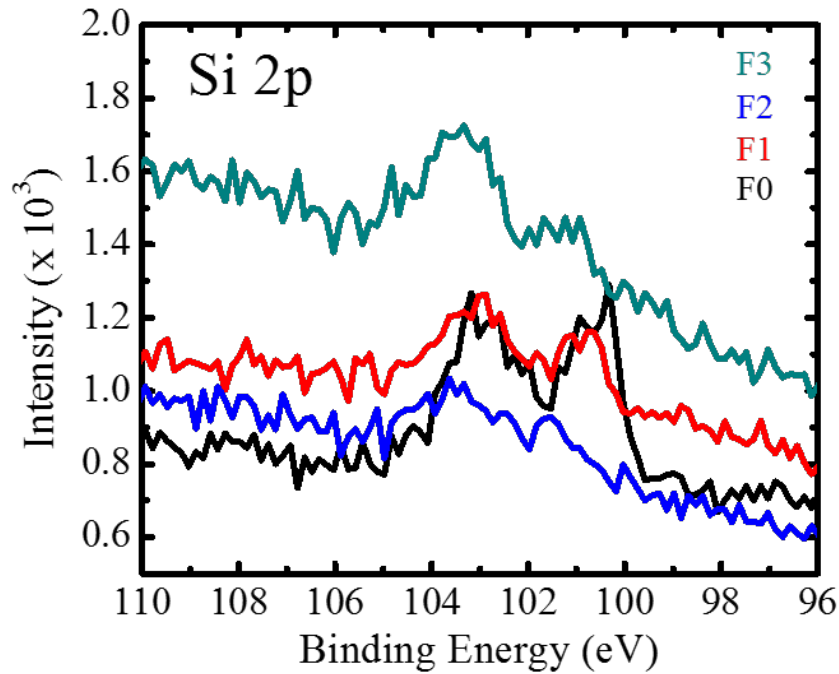

Figure S3: High resolution Si 2p spectra of the fluorinated device before, and after each cumulative fluorine dosage, labeled F0-F3

Upon annealing, the XPS spectra showed the recovery of the carbon peak to that of sp$^2$ graphene. The XPS C1s spectra showing the chemical recovery of the functionalized graphene device after vacuum annealing (conditions described for O2 and O3 in table 1) is shown in figure S4. The data show that after the second round of oxygen functionalization, the graphene device contained incorporated oxygen functionalities assigned to carbon bonding involving ethers or alcohols (C-O-C, C-O, or C-OH) and carbonyl bonds (=O) located at ≈286.4 eV and ≈287.1eV, respectively. After annealing (O3), the EG C-C sp$^2$ peak intensity increased tremendously, indicating restoration of the graphene lattice upon removal of the oxygen functional groups, clearly showing the partial removal of the attached functional groups and maintained integrity of the graphene sp$^2$ nature.

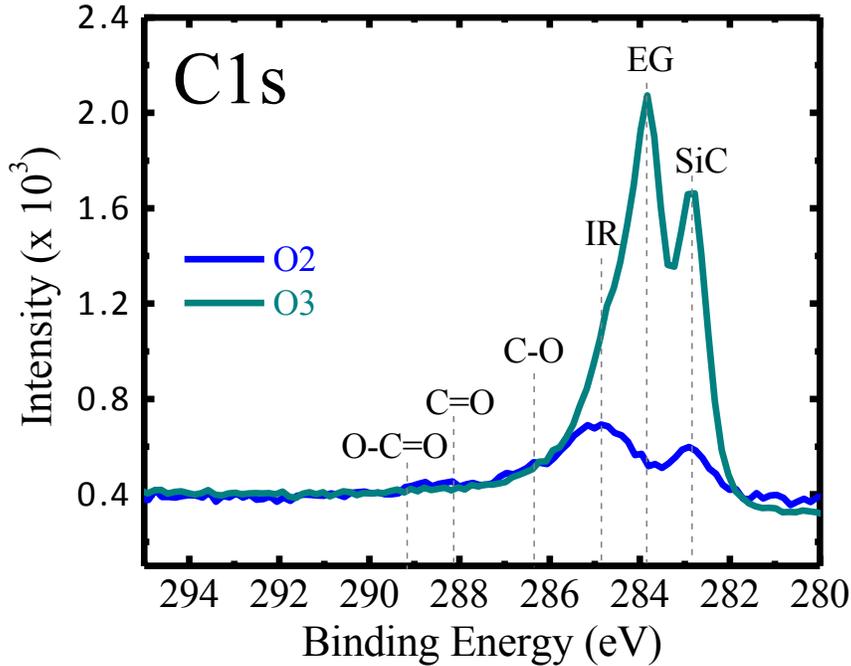

Figure S4: High resolution C1s spectra of an oxygenated device before, O2, and after, O3, vacuum annealing (450°C, 1hr). The graphene peak with $sp^2$ bonding is labeled as EG, the interfacial layer is labeled as IR, and the silicon carbide signal is referred to as SiC.

Raman characterization was performed using an InVia Raman microscope (Renishaw) equipped with a 50x objective lens. A 514.5nm diode laser provided the excitation with the scattered light dispersed by a 1800-line grating into a cooled detector array. Experiments were performed at a set power of 20mW at the source with a spot size of 5μm. Raman spectroscopy is powerful for identifying the number of layers, level of disorder, and doping of graphene. The Raman spectra of pristine and nitrogen functionalized EG devices is shown in Figure S5. Before functionalization there was a weak G peak ($\approx 1600$ cm$^{-1}$) and a 2D peak ($\approx 2730$ cm$^{-1}$), with no detectable D peaks (indicative of EG disorder) for all EG samples. Conjugation of the six member ring structure of graphene can become disrupted when functional groups are introduced to the carbon structure due to electron sharing or $sp^3$-bond formation. Notably, the conversion of the $sp^2$ carbon hybridization to $sp^3$ hybridization breaks symmetry, causing the activation of a breathing mode of the six-membered $sp^2$-carbon rings, which gives rise to a "disorder-induced" peak observed at $\sim 1340$ cm$^{-1}$ (D peak) in the Raman spectrum. Therefore, the presence of a D peak in the exposed region is indicative of the localization of defects induced by the incorporation of functional groups in those areas. After plasma treatment, the Raman spectra of the same device showed an increase in the disorder induced D peak and decreases in both the G and 2D peaks, which is characteristic of the disruption of $sp^2$ bonding of the graphene lattice. The intensity of the D line increased further at higher dosages due to the increased defect sites and decreased $sp^2$ cluster size. However, the 2D peak was still evident even at the highest dosage of functionalization, demonstrating the presence of graphene.

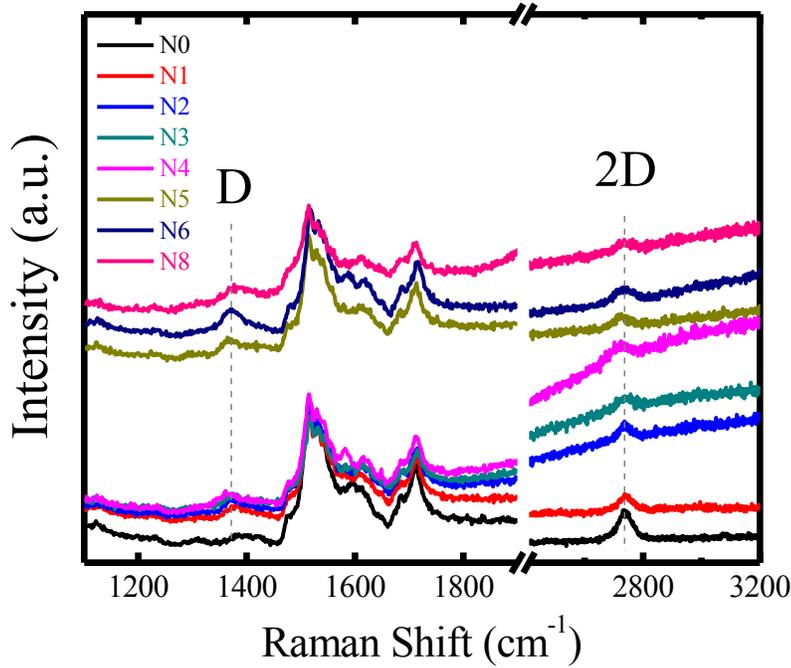

Figure: S5. Raman spectra of epitaxial graphene devices before and after nitrogen functionalization (N0 – N6), and after annealing.

Surface morphology was characterized by Atomic force microscopy (AFM, Bruker Dimension Icon) and scanning electron microscopy (SEM, Carl Zeis).

6. *Influence of plasma functionalization on transport properties.* Functionalization through plasma exposure produces scattering sites on the graphene that profoundly affects the resistance. The room temperature resistance rapidly increased with dose and saturated at approximately 8000Ω/square independent of the chemical nature of the dose (Fig. S6(a)). Hall measurements indicate that this increase is predominately due to decreased mobility although there is large scatter in the data for carrier concentration (Fig. S6(b) and (c)). The scatter may be due to variations in the position of the Dirac point which we are unable to determine in this study.

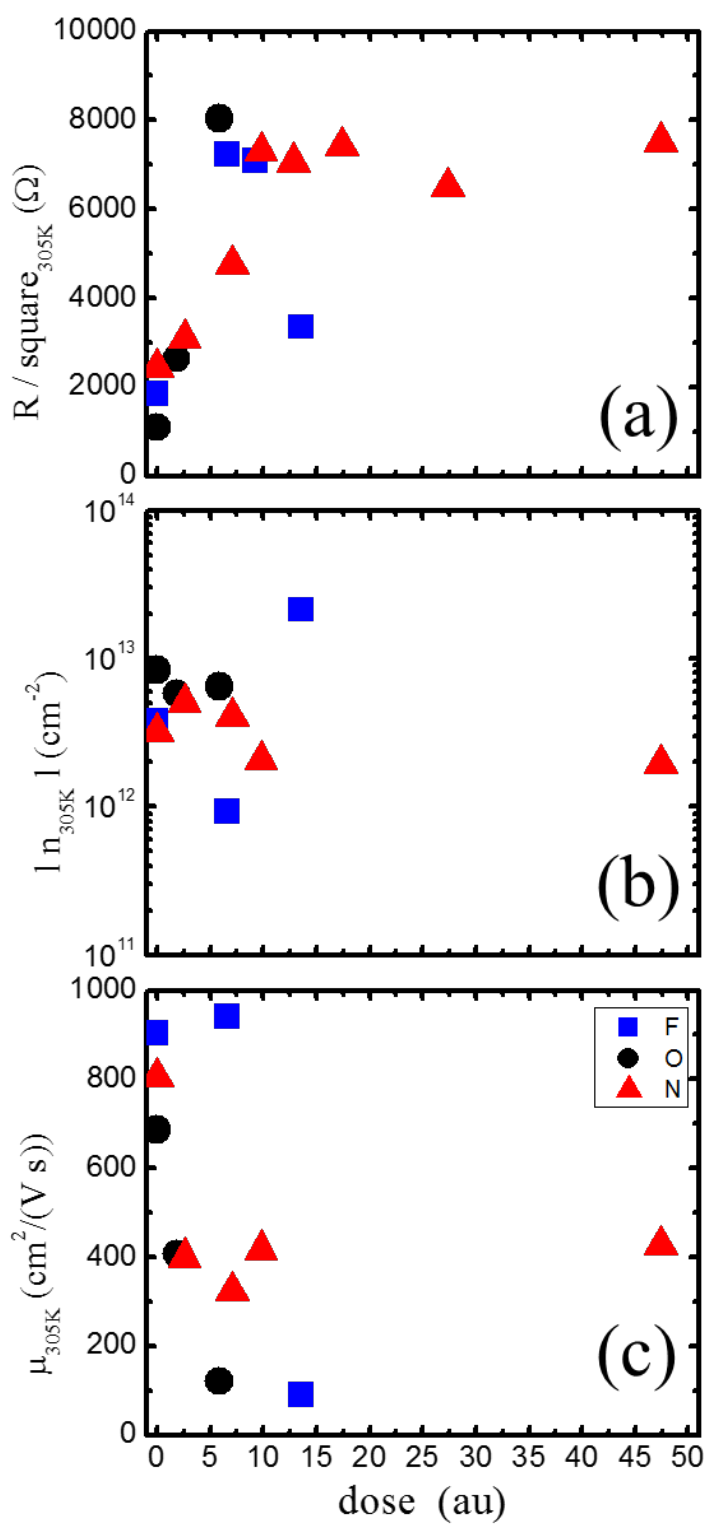

Figure S6. Room temperature transport properties as a function of plasma exposure: (a) resistance, (b) carrier density, and (c) mobility. F, O, and N refer to operating backgrounds containing $SF_6$, $O_2$, and $N_2$,

respectively. The horizontal axis is the same in all cases and can be seen in (c). The results exclude samples subject to vacuum anneals since heating will remove functional groups introduced during plasma exposure.

7.  *Heating effects.* It is well known that in conducting systems at low temperatures the energy transfer from electrons to the lattice can be limited by an electron-phonon bottleneck [*8,9,10,11,12,13*]. This phenomenon causes the decoupling of the electrons from the surrounding lattice when a measurement current is applied resulting in a non-equilibrium situation where the electrons have an effective temperature that exceeds that of the surrounding phonon bath. Experimentally, this steady-state manifests as a constant resistivity at low temperature because the electron temperature no longer tracks the measured ambient phonon temperature. This behavior is seen in Fig. S7 where the resistivity of a single sample obtained for the low measurement currents follows a logT dependence while that for the higher measurement currents saturate at values that decrease with increasing measurement current at low temperatures, consistent with the expectation that higher input power results in hotter carriers. The importance of this effect in graphene was noted by Baker, *et al*. [*14*] This result appears similar to that expected for the Kondo effect which has been reported in graphene with defects [*15*]. Also, Price, *et al*. [*16*] reported heating effects below 70K in exfoliated graphene flakes that were attributed to an electron-phonon bottleneck at the graphene substrate interface.

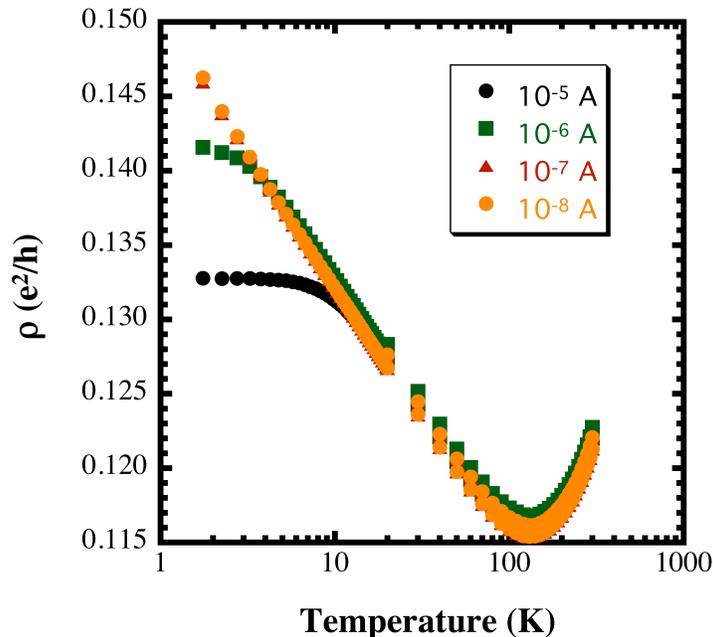

Figure S7. Sheet resistance vs. Temperature using several measurement currents for a sample exposed to $N_2$-containing plasma (sample N1). Note that saturation of sheet resistance occurs with higher measurement currents.

8.  *Temperature dependence of the Hall resistance.* Hall resistance data were obtained using a standard Hall bar geometry for -8T≤B≤8T in a Quantum Design Physical Property Measurement System (PPMS). The data for negative B were subtracted from those for positive B and averaged to remove the contribution from the magneto-resistance. The low temperature $R_{Hall}$ values for several plasma doses (see Table S1 for dose details; 0 indicates unexposed samples) are normalized to the $R_{Hall}$ values at 1.75K and are plotted as a function of log(T) in Fig. S8. These results demonstrated that like $R_{square}$, $R_{Hall} \sim \log T$.

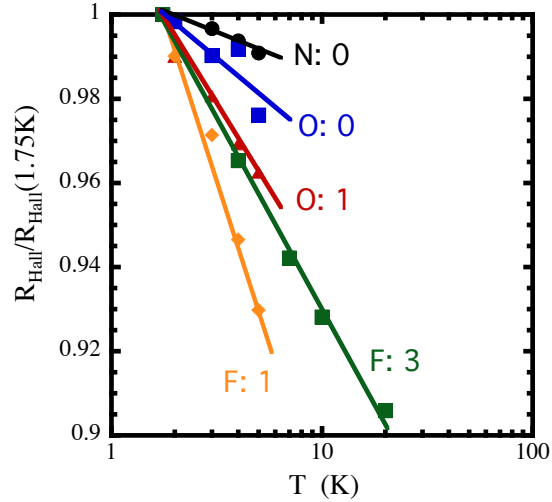

Figure S8. Normalized Hall coefficient vs. log(T) for several samples and plasma doses at low temperature. The lines are guides for the eye. The exposures are shown in table S1.

9.  *Relaxation times.* The decoherence (a), intervalley (b), and multicomponent (c) relaxation times extracted from fits of equation 3 in the main text to the magneto-resistance (MR) data are shown as a function of $\sigma(1K)$, a measure of the distance from the strongly localized phase are shown in Fig. S9. It is important to note that the theory was developed for systems far from the strongly localized state while these samples were very close and that one should be wary of taking the values of these times too seriously (see main text).

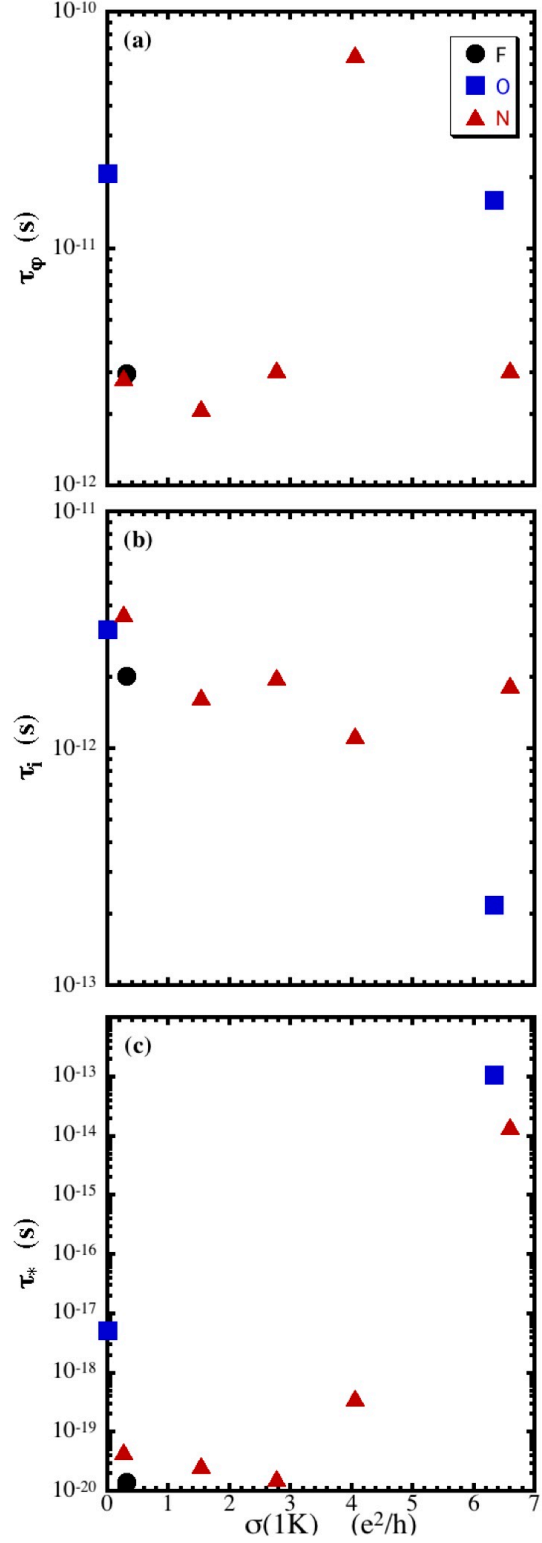

Figure S9. Characteristic time scales extracted from the MR data: (a) decoherence, (b) intervalley, and (c) $\tau_*$, defined as $\tau_*^{-1} = \tau_w^{-1} + \tau_z^{-1} + \tau_i^{-1}$, where $\tau_w$ is the relaxation time associated with the warping term and

$\tau_z$ is the intravalley relaxation time. The horizontal axis for (a), (b) and (c) are the same and is found in (c).